\documentclass[twocolumn,nopacs,preprintnumbers,amsmath,amssymb]{revtex4}

\usepackage{verbatim}
\usepackage{graphicx}% Include figure files
\usepackage{dcolumn}% Align table columns on decimal point
\usepackage{bm}% bold math

\begin{document}

\title{Raman spectroscopy and field emission measurements on catalytically grown carbon nanotubes}
\author{Christian Klinke}
\email{christian@klinke.org}
\author{Ralph Kurt}
\altaffiliation{Present address: Philips Research Laboratories
Eindhoven, Prof. Holstlaan 4 (WA 03), 5656 AA Eindhoven, The
Netherlands.}
\author{Jean-Marc Bonard}
\altaffiliation{Present address: Rolex S.A., 3-7 Rue Francois-Dussaud, 1211 Geneva 24, Switzerland.}
\affiliation{Institut de Physique des Nanostructures, Ecole Polytechnique F\'ed\'erale de Lausanne,\\
CH - 1015 Lausanne, Switzerland}

\author{Klaus Kern}
\affiliation{Institut de Physique des Nanostructures, Ecole Polytechnique F\'ed\'erale de Lausanne,\\
CH - 1015 Lausanne, Switzerland \\ \textnormal{and} \\ Max-Planck-Institut f\"ur Festk\"orperforschung, D - 70569 Stuttgart, Germany}

\begin{abstract}

We used microcontact printing to pattern a silicon surface with an
iron-containing catalytic solution. Multi-wall carbon nanotubes
were subsequently grown on the patterned areas by chemical vapor
deposition at temperatures between 650 and 1000${^\circ}$C. We
demonstrate that the diameter of the catalytically grown
multi-wall nanotubes increases with the deposition temperature.
Raman spectroscopy has been used to investigate the crystalline
character of the obtained structures and it is found that the
fraction of the nano-crystalline shell increases with the
temperatures. The measurement of the field emission properties
shows a correlation between the tube diameter and the emission
field values.

\end{abstract}

\maketitle

\section*{Introduction}

Carbon nanotubes~\cite{IIJIMA} have been studied already for
several years and are now considered for applications in
miscellaneous devices such as tubular lamps~\cite{CHATELAIN}, flat
panel displays~\cite{KIM}, lighting elements~\cite{YAMAKAWA} or
nanometric electronic devices~\cite{DEKKER,YAMAKAWA2}. Such
devices make precise demands on the properties of the tubes, as
the length, diameter and electronic properties have a strong
influence on the final performance of the device. This implies
that the nanotube growth has to be controlled and understood.
However, we still lack an exact knowledge of the growth mechanism,
be it for arc discharge, laser ablation, or catalytic
growth~\cite{IIJIMA2}.

One of the most promising properties which should be exploited for
applications is the very good field emission of the
nanotubes~\cite{WEISS}. The catalytic deposition is the most
practical method to create nanotubes on vast surfaces. In a recent
paper we studied the deposition conditions of the catalytic growth
of multiwalled carbon nanotubes~\cite{KERN}. We found that the
catalyst had an influence on the morphology of the grown
structures, and that Fe was better suited for the low temperature
growth of nanotubes than Ni or Co. We also noted that the diameter
of the nanotubes increases with the deposition temperature. We
present here advanced investigations on the structures obtained
with Fe as catalyst. We performed Raman and Transmission Electron
Microscopy (TEM) studies to elucidate the crystalline structure of
these objects. In respect of applications we also completed the
research on these structures with the study of their field
emission properties. We demonstrate that combing field emission
measurements, Scanning Electron Microscopy (SEM), Raman
spectroscopy and TEM allows an optimization of crystallinity,
geometry and field emission while gaining new information about
the structure of the considered objects.

\section*{Experimental methods}

\subsection*{Synthesis of nano-structured material}

Microcontact printing~($\mu$CP) was performed to pattern a
$<$100$>$-oriented boron doped silicon wafer (resistivity: 5 - 25
m$\Omega\cdot$cm)~\cite{KERN}. The stamps for $\mu$CP were
obtained by curing poly\-(dimethyl)\-siloxane~(PDMS) for at least
12~h at 60$^{\circ}$C on a structured master which was prepared by
contact photolithography. The structures of these stamps are
squares with a width of 5~$\mu$m.  They were hydrophilized before
use by a oxygen plasma treatment (O$_{2}$ pressure $\sim
$0.8~mbar, load coil power $\sim $75~W, 60~s). The catalyst
solution for the $\mu$CP was a 100~mM solution of
Fe(NO$_{3}$)$_{3}\cdot$9H$_{2}$O in ethanol.  The catalyst
concentration of 100~mM was chosen since at this amount a dense,
but well-separated growth of nanotubes is obtained with good
reproducibility~\cite{KERN}. For printing the stamp was loaded
with 0.2~ml of catalyst solution for 30~s and then dried in a
nitrogen stream for 10~s. The printing was performed by placing
the stamp on the surface of the SiO$_{2}$/Si wafer for 3~s.

The catalytic growth of nanotubes was carried out in a flow
reactor (quartz tube with an inner diameter of 14~mm in a
horizontal oven) directly after the printing. Before the
deposition the volume of the quartz tube was rinsed by a nitrogen
stream of 80~ml/min. The deposition was performed with 80~ml/min
of nitrogen and 20~ml/min of acetylene (carbon source for the
catalytic growth) at atmospheric pressure.

\subsection*{Characterization techniques}

Scanning electron microscopy~(SEM) was performed to analyze the
microstructures in plan view. A Philips XL~30 microscope equipped
with a field emission gun~(FEG) was used with an acceleration
voltage between 3 and 5~kV, a working distance of 10~mm, and in
secondary electron~(SE) image mode.

The growth morphology and crystallinity of the tubular structures
were controlled by transmission electron microscopy~(TEM). For
this purpose a Philips EM~430 microscope equipped with a Gatan
image plate operating at 300~kV (point resolution 0.3~nm) was
used.

Information about the vibrational properties of the nanostructures
was obtained by micro-Raman spectroscopy. The Raman spectra were
recorded in backscattering configuration using the 514.5~nm line
of an Ar$^{+}$ ion laser and a DILOR XY~800 spectrometer. An
incident maximum laser power of 20~mW was applied in order to
avoid peak shifts due to thermal heating or structure
transformations during data acquisition. A spot size of
approximately 2~$\mu$m was achieved with a 250$\times$ Olympus
microscope objective. The spectra were calibrated using a natural
diamond single crystal.

The field emission measurements were performed using the examined
samples as cathodes. The emitted electrons were collected on a
highly polished stainless steel spherical counterelectrode of 1~cm
diameter, which corresponds to an emission area of
$\sim$0.007~cm$^{2}$. The distance between the electrodes was
adjusted to 125~$\mu$m. A Keithley~237 source-measure unit was
used to supply the voltage (up to 1000~V) and to measure the
current with pA sensitivity, allowing the characterization of
current-voltage ($I-V$) behaviour.

\section*{Results}

\subsection*{Morphology}

\begin{figure}[!h]
\begin{center}
\includegraphics[width=0.45\textwidth]{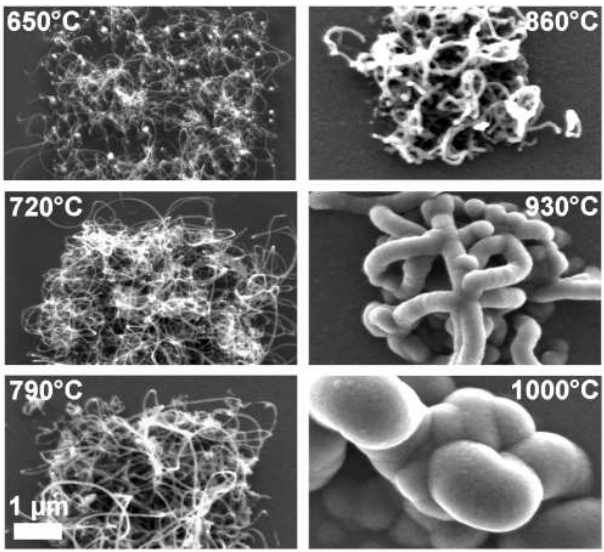}
\caption {\it SEM micrographs of carbon structures obtained at
deposition temperatures from 650${^\circ}$C up to 1000${^\circ}$C
(deposition time: 30~min).} \label{TEMPERATURE}
\end{center}
\end{figure}

\begin{figure}[!h]
\begin{center}
\includegraphics[width=0.45\textwidth]{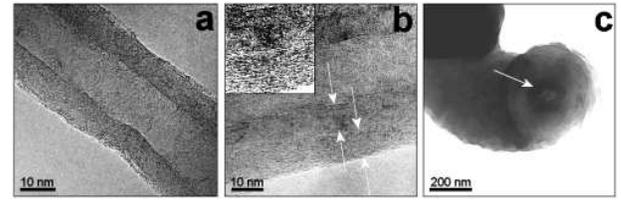}
\caption {\it (a) TEM micrograph of a section of a carbon nanotube
grown at 650${^\circ}$C. The graphitic layers of the hollow
multi-wall nanotube are well visible. (b) TEM micrograph of a
carbon structure grown at 790${^\circ}$C with a crystalline core
and a polycrystalline outer shell. The core and the shell are
delimited by arrows. The region of interest is enlarged in the
inset. (c) TEM micrograph of a carbon structure grown at
930${^\circ}$C with an arrow is indicating the nanotube core.}
\label{TEM}
\end{center}
\end{figure}

The influence of the deposition temperature on the morphology of
the carbon structures is demonstrated for a deposition time of
30~min on Fig.~\ref{TEMPERATURE}. The carbon nanotubes and carbon
structures grow with random orientation from iron-inked squares of
a silicon surface. One can nicely see the increase of the diameter
from thin nanotubes at 650${^\circ}$C to thick ``carbon worms'' at
1000${^\circ}$C. The diameter varies between 25~nm for the
structures at 650${^\circ}$C and about 1~$\mu$m for the structures
at 1000${^\circ}$C. TEM reveals that multi-wall nanotubes grown at
650${^\circ}$C are hollow and well-graphitized (Fig.~\ref{TEM}a) .
The well-separated nanotubes have an inner diameter of about 15~nm
and an outer diameter of about 30~nm. Most of them have open ends
and some nanotubes contain encapsulated catalyst particles. In
about ten percent of the nanotubes we found these particles at the
top of the tube. In this case, the particles are of prolate shape
and aligned in the growth direction~\cite{KERN}.

The carbon structures grown at higher temperatures consist of a
``nanotube core'' and an additional layer of amorphous or
polycrystalline carbon. Fig.~\ref{TEM}b shows a TEM image of the
structures obtained at 930${^\circ}$C where a core structure is
surrounded by flake-like carbon (indicated by arrows in
Fig.~\ref{TEM}b). The structures at 1000${^\circ}$C are too thick
to be imaged by TEM due to the electron transparency.

\begin{figure}[!h]
\begin{center}
\includegraphics[width=0.45\textwidth]{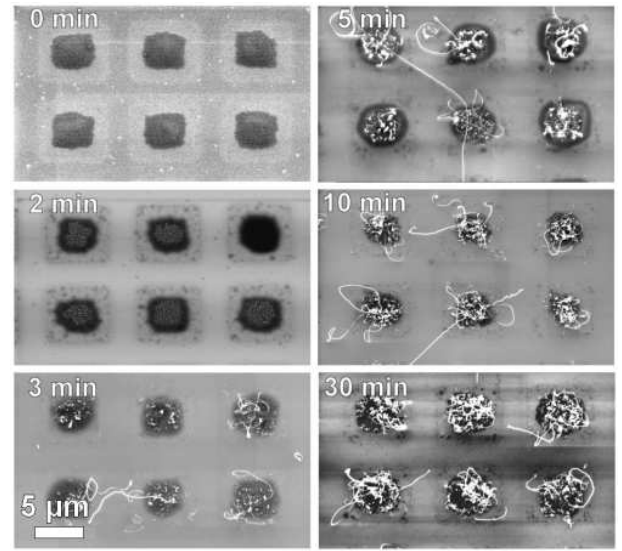}
\caption {\it Time dependence of the growth of the carbon
nanotubes by chemical vapor deposition (deposition temperature:
720${^\circ}$C).} \label{TIME}
\end{center}
\end{figure}

We also found that the growth of the nanotubes is a very fast
process: under our conditions the growth takes place during the
first 5~min already. Fig.~\ref{TIME} shows the evolution of the
nanotube growth with time at 720${^\circ}$C. After the annealing
without the CVD process only the printed catalyst patterns are
visible. After 2~min of deposition some dots of carbonized
catalyst appear in the center of the printed squares. Only 1~min
later, we detect nanotubes of up to 10~$\mu$m length, which
implies that the growth rate is of at least 160~nm/s. In the
following time the density of the nanotubes increases but we
cannot detect a significant increase in length. The maximum length
remains at about 10~$\mu$m.

\subsection*{Raman spectroscopy}

\begin{figure}[!h]
\begin{center}
\includegraphics[width=0.45\textwidth]{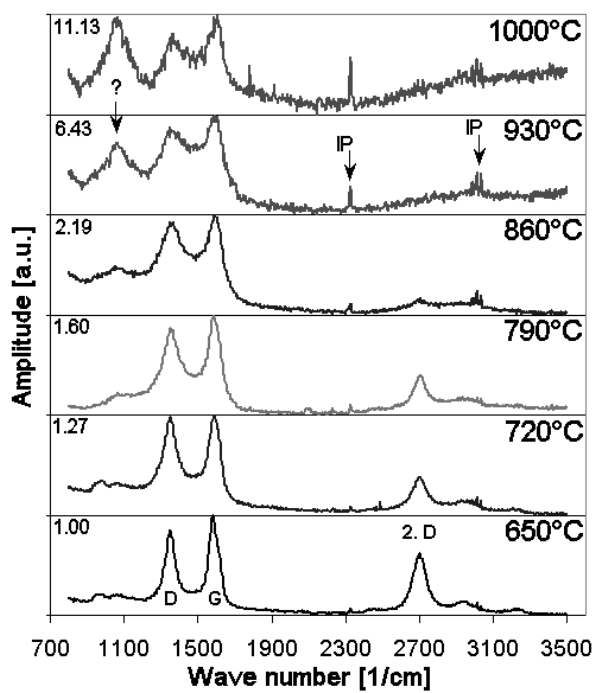}
\caption {\it Micro-Raman spectroscopy of carbon nanotubes
obtained at temperatures between 650${^\circ}$C and
1000${^\circ}$C. The peaks for the disordered~(D) and the
graphitic~(G) carbon, the second order D-peak (2.D) and impurity
peaks~(IP) are visible. On the left side of the spectra the
scaling factors are mentioned (relative to the spectrum at
650${^\circ}$C). The spectra are normalized to the highest peak
(G-peak) in each spectrum. } \label{RAMAN}
\end{center}
\end{figure}

We performed micro-Raman spectroscopy in order to investigate the
vibrational properties of the synthesized carbon structures, which
allows also to draw further conclusions about their
crystallography or morphology. Fig.~\ref{RAMAN} compares the Raman
spectra measured from carbon nanotubes grown at temperatures
between 650${^\circ}$C and 1000${^\circ}$C. All spectra show at
least the two significant peaks at 1580~cm$^{-1}$ and at
1347~cm$^{-1}$, which become broader at higher temperatures and
overlap.

Crystalline graphite leads to a sharp vibration mode at
1580~cm$^{-1}$~\cite{KOENIG} which is due to the presence of
C-sp$^{2}$ domains and named first order G band. The peak at
approximately 1350~cm$^{-1}$ is considered to represent a more
disordered structure and is labeled as D (disordered)
band~\cite{KATKANANT}. Note that in a perfect graphite crystal the
first order vibrational mode of the D band is forbidden due to the
selection rules. Decreasing particle size or bending of the
lattice fringes may activate this band. As seen in
Fig.~\ref{RAMAN}, the second order D peak (2.D) appears at
approximately 2700~cm$^{-1}$ for nanotubes grown at lower
temperatures. However, with increasing deposition temperature this
peak disappears. The spectra are normalized to the highest peak in
each spectrum (the G-peak). The signal strength gets weaker for
the structures deposited at higher temperatures, therefore the
noise level becomes more and more visible in the spectra.

It is known (e.g.~\cite{WITHE,ANDERSON}) that smaller particles as
well as structural imperfections will broaden the first order
peaks from graphite. Therefore one can estimate the order of
crystallinity in the material from the corresponding halfwidth
(FWHM). An amorphous structure leads typically to a halfwidth
(FWHM) of approximately 200~cm$^{-1}$~\cite{ANDERSON} as observed
in the case of deposition at 1000${^\circ}$C. In the case of
nanotubes deposited at 650${^\circ}$C sharp peaks (FWHM
$\sim$90~cm$^{-1}$) reveal their much higher degree of crystalline
perfection.

The catalytically grown carbon nanotubes can be characterized as a
nano-crystalline but disordered graphite-like system where the
disorder increases with the preparation temperature. This confirms
qualitatively the results obtained by TEM. Unfortunately, Raman
measurements could not clearly confirm that the high temperature
carbon structures consist of an amorphous and of a crystalline
part as suggested by TEM.

Interestingly, the relative height of the peak at about
1047~cm$^{-1}$ increases with temperature. This peak could not yet
be identified, but its broad shape indicates that it could
probably originate from solid state phonons. We already detected
this peak in earlier experiments on nitrogenated carbon
nanotubes~\cite{KARIMI}. Some other peaks as indicated by arrows
in Fig.~\ref{RAMAN} might be due to impurities. Stretch vibrations
of N$_{2}$ in the ambient air might cause the sharp peak at about
2325~cm$^{-1}$~\cite{WOTTKA}.

\subsection*{Field emission}

\begin{figure}[!h]
\begin{center}
\includegraphics[width=0.45\textwidth]{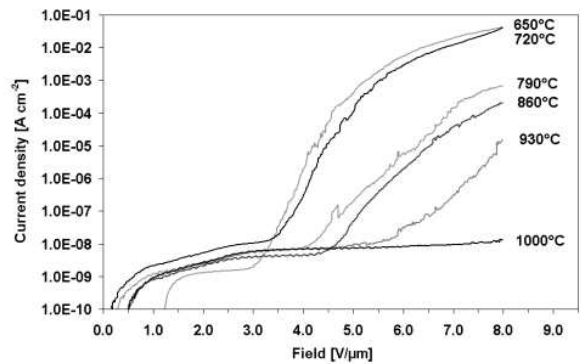}
\caption {\it Field emission of the carbon nanotubes. The
nanotubes obtained at 650${^\circ}$C provide the lowest emission
field values.} \label{FIELDEMISSION}
\end{center}
\end{figure}

In Fig.~\ref{FIELDEMISSION} we present the results of the field
emission measurements of the obtained carbon films. We noted a
decrease of the absolute current density at given applied field
with increasing deposition temperature. The turn-on field~$E_{\rm
to}$ (field to obtain a current density of 10$^{-5}$~A/cm$^{2}$,
first illumination of a screen pixel) and the threshold field
$E_{\rm thr}$ (field at a current density of 10$^{-2}$~A/cm$^{2}$,
saturation of a screen pixel) both increase with increasing
deposition temperature~(Tab.~\ref{FE-TABLE}). For the carbon
structures deposited at temperatures of 790${^\circ}$C and above
the current density for the threshold field of
10$^{-2}$~A/cm$^{2}$ was not reached below the maximal applied
voltage of 1000~V at 125~$\mu$m interelectrode distance.

The field amplification factor was calculated with the
Fowler-Nordheim formula. The model describes the electron emission
from a flat surface by tunneling through the triangular surface
potential barrier. The emitted current $I$ is proportional to
$F^{2} \exp (B \phi^{3/2} / F)$, where $F$ is the applied field
just above the emitting surface, $\phi$ is the work function and
$B$ is a constant ($B = 6.83\cdot 10^{-9}$V eV$^{3/2}$
m$^{-1}$)~\cite{PLUMMER}. Generally, $F$ is not known exactly and
is therefore taken here as $F = \beta E = \beta V/ d$, with the
applied voltage $V$, the interelectrode distance $d$ and the
macroscopic applied field $E = V / d$. The work function was
assumed to be equal to 5~eV, which is a reasonable assumption for
carbon-based field emitters~\cite{SCHLAPBACH}.

The field amplification values do not follow a simple trend with
temperature, as can be extracted from Tab.~\ref{FE-TABLE}. It is
well-known that for a single tube, a larger diameter will lower
the field amplification factor for a given length. The increase of
the diameter found in Fig.~\ref{TEMPERATURE} should thus result in
a monotonously decreasing field amplification factor. Our
experimental results suggest that this trend is masked (at least
in part) by varying nanotube lengths and nanotube densities on the
samples, which lead to more or less pronounced screening
effects~\cite{CHATELAIN2}. The comparison with the SEM images
suggest that the field emission corresponds to the geometry of the
structures and that the structures with a smaller diameter emit
better. This behavior was also observed earlier for nitrogenated
carbon structures~\cite{KLINKE}.

\section*{Discussion}

\begin{figure}[!h]
\begin{center}
\includegraphics[width=0.45\textwidth]{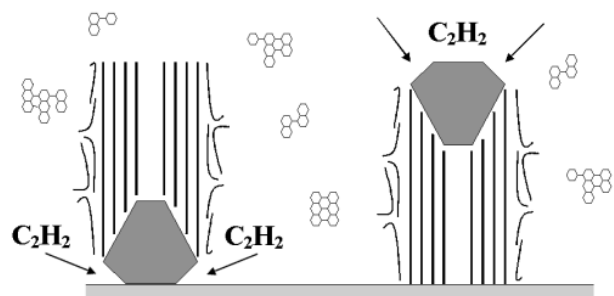}
\caption {\it Model for the supposed growth mechanism of
catalytically grown carbon nanotubes (based on considerations by
Kanzow et al.~\cite{DING}).} \label{MODEL}
\end{center}
\end{figure}

We discuss in the following our results in the light of the most
probable growth mechanism for carbon nanotubes under our
experimental conditions. Acetylene is stable at temperatures below
800$^{\circ}$C and can be only catalytically dissociated, in our
case on the small metal (oxide) particles delivered to the
substrate by microcontact printing (Fig.~\ref{MODEL}). The
dissociation reaction takes presumably place at facets of
well-defined crystallographic orientation, and the resulting
hydrogen H$_{2}$ is removed by the gas flow whereas the carbon is
dissolved in and diffuses into the particle~\cite{DING}. For
unsaturated hydrocarbons this process is highly exothermic. When
the particle is saturated with carbon, the carbon segregates on
another, less reactive surface of the particle, which is an
endothermic process. The resulting density gradient of carbon
dissolve in the particle supports the diffusion of carbon through
the particle. To avoid dangling bonds, the carbon atoms assemble
in a sp$^2$ structure at a less reactive facet of the particle,
which leads to the formation of a nanotube.

We noted in Fig.~\ref{TIME} that the nanotube growth did not begin
immediately after the introduction of the hydrocarbon gas in the
reactor, but that some carborized spots appear before the rapid
nanotube growth. This suggests that a certain quantity of carbon
must be dissolved into and diffuse through the particle before the
nanotube growth can start.

The simple model presented in Fig.~\ref{MODEL} describes the
growth with a particle at the top of the nanotube or at the
bottom. In the second case the particle sticks more to the
substrate surface than in the first case. But there must be free
particle surfaces which are exposed to the gas to proceed the
growth. In the second case the acetylene diffuses from the side
into the particle and the nanotube is constructed from the bottom
up, whereas in the first case the gas diffuses from the sides and
form the top into the particle. This seems to be the favored
mechanism in our case as typically 90~\% of the tubes have closed
tips without a catalytic particle.

It is at first glance difficult to understand why the diameter of
the structures increases with the temperature above 800$^{\circ}$C
in the frame of the above model. At around 800${^\circ}$C,
acetylene starts to dissociate spontaneously, and the reaction gas
contains therefore a significant fraction of free carbon, which
will form larger aggregates in order to avoid dangling bonds.
These carbon flakes, once formed, are carried with the gas flow
and may be deposited on the substrate. We therefore propose the
following scenario at temperatures above 800${^\circ}$C: as is the
case at lower temperatures, carbon nanotubes of small diameter are
formed over the catalyst patterned areas after an activation
period, while the growth itself takes place very rapidly. In
addition, the flakes formed in the gas phase condensate on the
substrate and on the formed nanotubes, adding a polycrystalline
outer shell over the graphitic inner core. The structures get
thicker with temperature because the proportion between
dissociated and molecular acetylene in the gas phase increases.

This explanation is supported by the TEM images and the Raman
spectroscopy, which confirmed the polycrystalline character of
these structures. In fact, while SEM suggests an amorphous carbon
structure on the surface of the grown tubes, TEM reveals a
crystalline core structure which is surrounded by a
polycrystalline shell (Fig.~\ref{TEM}b and c). Raman spectroscopy
showed that there is a continuous increase of the polycrystalline
fraction in the structures what corroborates the findings. The
outer shell becomes thicker with higher temperature starting from
the thinnest structures at a temperature of 650${^\circ}$C.

The results of the field emission experiments show that the
thinnest nanotubes are more efficient field emitters. The field
emission properties (emission fields, field amplification factor)
follow loosely the morphology of the individual tubes, as the
emission fields decrease with increasing temperature. It seems
however that the overall structure of the nanotube film, such as
nanotube density and height, plays a role that is more important
than the diameter of the structures because of screening effects.

The crystallinity of the nanotubes may also influence the field
emission properties. As the work function of polycrystalline and
graphitic carbon are very similar, the main difference between the
two forms is their electrical resistivity, which is lower in the
case of well-graphitized carbon. A higher resistivity will lead to
higher emission fields as a voltage drop will appear along the
tube, reducing the effective applied field. However, this effect
will play a role only at high current densities, and that it will
have little influence on the low current part of the I-V curve.
Therefore significant differences in turn-on field and field
enhancement factor between well-graphitized and polycrystalline
nanotubes of equivalent dimensions are not expected. The major
factor that determines the field emission properties is the
nanotube diameter, length and spacing.

\section*{Conclusions}

We have grown carbon nanostructures by thermal CVD with an iron
catalyst that was delivered to the Si substrate by microcontact
printing. We noted that the diameter and morphology of the
produced structures varied with the deposition temperature, from
thin and well-graphitized carbon nanotubes at 650${^\circ}$C to
$\mu$m-thick fibers at 1000${^\circ}$C. We used TEM imaging, Raman
spectroscopy and field emission experiments to investigate in more
detail the character of these structures, and found that the
increase in temperature above 800${^\circ}$C resulted in the
formation of a polycrystalline outer shell over a nanotube core.
We suggest that this effect is due to the dissociation of
acetylene in the gas phase, which leads to the formation of carbon
flakes that are subsequently deposited on the catalytically grown
structures.

\section*{Acknowledgement}

The Swiss National Science Foundation~(SNF) is acknowledged for
the financial support. The electron microscopy was performed at
the Centre Interd\'epartemental de Microscopie Electronique~(CIME)
of EPFL. The authors are grateful to Heiko Seehofer from
D\'epartement des Materiaux of EPFL for technical assistance in
Raman spectroscopy.

% \newpage

\section*{Figures \& Tables}

\begin{table}[!b]
\begin{center}
\caption {\it Table of the field emission values: The turn-on
field~E$_{\rm to}$ (field at a current density of
10$^{-5}$~A/cm$^{2}$, first illumination of a screen pixel), the
threshold field~E$_{\rm thr}$ (field at a current density of
10$^{-2}$~A/cm$^{2}$, saturation of a pixel) and the field
amplification~$\beta$ (obtained by calculations based on the
Fowler-Nordheim theory) as function of the deposition
temperature.} \label{FE-TABLE} \vspace{0.5cm}
\begin{tabular}{|p{1.8cm}|p{2.0cm}|p{2.2cm}|p{1.8cm}|}
  \hline
  Temperature [${^\circ}$C]& Turn-on field $E_{\rm to}$ [V/$\mu$m] & Threshold
  field $E_{\rm thr}$ [V/$\mu$m] & Field amplification $\beta$\\ \hline
  650 & 4.2 & 6.4 & 696 \\ \hline
  720 & 4.5 & 6.8 & 688 \\ \hline
  790 & 6.3 & - & 958 \\ \hline
  860 & 6.5 & - & 751 \\ \hline
  930 & 7.9 & - & 410 \\ \hline
  1000 & - & - & - \\ \hline
\end{tabular}
\end{center}
\end{table}

\clearpage


\begin{thebibliography}{00}

\bibitem{IIJIMA}
Iijima, S. {\it Nature} {\bf 1991}, 354, 56.

\bibitem{CHATELAIN}
Bonard, J. M.; St\"ockli, T.; Noury, O.; Ch\^atelain, A. {\it
Appl.  Phys.  Lett.} {\bf 2001}, 78, 2775.

\bibitem{KIM}
Choi, W. B.; Chung, D. S.; Kang, J. H.; Kim, H. Y.; Jin, Y. W.;
Han, I. T.; Lee, Y. H.; Jung, J. E.; Lee, N. S.; Park, G. S.; Kim,
J. M. {\it Appl. Phys.  Lett.} {\bf 1999}, 75, 3129.

\bibitem{YAMAKAWA}
Murakami, H.; Hirakawa, M.; Tanaka, C.; Yamakawa, H. {\it Appl.
Phys. Lett.} {\bf 2000}, 76, 1778.

\bibitem{DEKKER}
Yao, Z.; Postma, H. W. C.; Balents, L.; Dekker, C. {\it Nature}
{\bf 1999}, 402, 273.

\bibitem{YAMAKAWA2}
Hirakawa, M.; Sonoda, S.; Tanaka, C.; Murakami, H.; Yamakawa, H.
{\it Appl. Surf.  Sci.} {\bf 2001}, 169, 662.

\bibitem{IIJIMA2}
Charlier, J. C.; Iijima, S. in: M. S. Dresselhaus, G. Dresselhaus,
P. Avouris (Eds.), Carbon Nanotubes (Springer, 2001).

\bibitem{WEISS}
Bonard, J. M.; Croci, M.; Klinke, C.; Kurt, R.; Noury, O.; Weiss,
N. {\it Carbon} {\bf 2002} in press.

\bibitem{KERN}
Klinke, C.; Bonard, J. M.; Kern, K. {\it Surf. Sci.} {\bf 2001},
492, 195.

\bibitem{KOENIG}
Tuinstra, F.; Koenig, J. L. {\it J. Chem. Phys.} {\bf 1970}, 53,
1126.

\bibitem{KATKANANT}
Dillon, R. O.; Woollam, J. A.; Katkanant, V. {\it Phys. Rev. B}
{\bf 1984}, 29, 3482.

\bibitem{WITHE}
Knight, D. S.; Withe, W. B. {\it J. Mater. Res.} {\bf 1989}, 4,
385.

\bibitem{ANDERSON}
Beeman, D.; Silverman, J.; Lynds, R.; Anderson, M. R. {\it Phys.
Rev. B} {\bf 1984}, 30, 870.

\bibitem{KARIMI}
Kurt, R.; Klinke, C.; Bonard, J. M.; Kern, K.; Karimi, A. {\it
Carbon} {\bf 2001}, 39, 2163.

\bibitem{WOTTKA}
Doerk, T.; Ehlbeck, J.; Jauernik, P.; Stancot, J.; Uhlenbusch, J.;
Wottka, T. {\it J. Phys. D} {\bf 1993}, 26, 1015.

\bibitem{PLUMMER}
Gadzuk, J. W.; Plummer, E. W. {\it Rev. Mod. Phys.} {\bf 1973},
45, 487.

\bibitem{SCHLAPBACH}
K\"uttel, O. M.; Gr\"oning, O.; Emmenegger, C.; Nilsson, L.;
Maillard, E.; Diederich, L.; Schlapbach, L. {\it Carbon} {\bf
1999}, 37, 745.

\bibitem{CHATELAIN2}
Bonard, J. M.; Weiss, N.; Kind, H.; St\"ockli, T.; Forro, L.;
Kern, K.; Ch\^atelain, A. {\it Adv.  Mater.} {\bf 2001}, 13, 184.

\bibitem{KLINKE}
Bonard, J. M.; Kurt, R.; Klinke, C. {\it Chem.  Phys.  Lett.} {\bf
2001}, 343, 21.

\bibitem{DING}
Kanzow, H.; Schmalz, A.; Ding, A. {\it Chem.  Phys.  Lett.} {\bf
1998}, 295, 525.

\end{thebibliography}
\end{document}